\documentclass{xeus}
\usepackage{graphics}
\usepackage{psfig}

\begin{document}
\onecolumn
\title{The Compton-thick AGN in the GPS radio source OQ+208}
\author{M.~Guainazzi\inst{1} \and C.~Stanghellini\inst{2} \and P.~Grandi
\inst{3}}
\institute{XMM-Newton Science Operation Center, RSSD, ESA, VILSPA, Apartado 50727, E-28080 Madrid, Spain
\and
Istituto di Radioastronomia del CNR- CP 141 , I-96017 Noto, Italy
\and
IAS/CNR - Via Fosso del Cavaliere 100, I-00133 Roma, Italy
}
\maketitle

\begin{abstract}

We report in this {\it paper} the ASCA discovery
of the first (to our knowledge) radio-loud
Active Galactic Nucleus (AGN)
covered by a Compton-thick X-ray absorber, in the GigaHertz
Peaked Spectrum radio source OQ+208. It represents one of
the few available direct  measurements of dense matter in the nuclear
environment of this class of sources, which may
provide the confining medium to the radio-emitting region if
GPS sources are indeed "frustrated" classical radio doubles.
The perspective of future studies  with XEUS are discussed.

\end{abstract}

\section{Introduction}

GigaHertz Peaked Spectrum radio sources are a class of powerful
($L_{radio} \sim 10^{45}$~erg~s$^{-1}$) radio sources, defined by a simple
convex spectrum peaking near 1 GHz. They represent $\simeq$10\% of the 5~GHz
selected sources, {\it i.e.} a significant fraction of the powerful radio
sources in the universe. They are characterized by
compact radio cores, most likely not extending beyond the
narrow-line regions (NLRs;
$\le$1 kpc). Some of them exhibit very faint extended radio
emission on scales larger than the host galaxy (Baum et al. 1990;
Stanghellini et al. 1990; Fanti et al. 2001), but rarely on Mpc scales
(Schoenmakers et al. 1999). Their radio morphologies and host galaxies
are generally consistent with
classical 3CR doubles. Emission-lines optical spectra
suggest interaction between the 
radio source and emission-line gas, as well as dust obscuration.
Mid-IR measurements suggest the existence of a powerful hidden
active nucleus
(Heckman et al. 1994: see, however,
de Vries et al. 1998 and Fanti et al. 2000
for a different point of view).

Two scenarios are currently proposed to explain the nature of GPS sources:

\begin{enumerate}

\item the {\it young scenario}: GPS could be young versions of large scale
radio galaxies at an earlier stage of their formation (Carvalho 1985:
Mutel \& Philips 1988; Fanti et al. 1995; Readhead et al. 1996: O'Dea \&
Baum 1997). In this scenario, objects which are $\sim$100 pc
in size are about
10$^4$~years old, which is consistent with proper motion measurements
on about 10 of them (Fanti 2000)

\item the {\it frustrated scenario}: GPS may represent "aborted" classical
doubles, which will never reach their full maturity because they are embedded
in a dense and turbulent medium, able to confine and trap the radio-emitting
region on the scale-length of the NLRs (van Brueghel et al. 1984;
O'Dea et al. 1991)

\end{enumerate}

X-ray observations
can provide an important contribution to elucidate
the nature of this class of objects. In facts:

\begin{itemize}

\item measurements of hot gas, through its optically thin emission peaking
in the soft X-rays, may provide indication for the presence of hot confining
medium (O'Dea et al. 1996) . Constraints on the presence of such a gas phase
from other wavelengths are not conclusive (Kameno et al. 2000; Marr et al. 2001)

\item measurements of heavy X-ray absorption (in the most extreme case
Compton-thick, $N_H > \sigma_T^{-1}
\simeq 10^{24}$~cm$^{-2}$) may indicate the presence of cold
confining matter, to be compared with hydrodynamical models
(De Young 1993; Carvalho 1994, 1998) to identify possible mechanisms
responsible for the confinement of the jet

\item the detection of large-scale X-ray jets may challenge one or more of our
current assumptions on the nature of GPS sources
(Siemiginowska et al. 2002)

\end{itemize}

The properties of the cold and hot phases,
that one can derive from the X-ray spectral fitting may therefore
provide a test for the "frustration" scenario.

Unfortunately, GPS sources are X-ray
weak: only 3 out of 9 GPS sources observed in
the soft X-rays have been detected. Hints that this low rate may be due to
absorption were put forward by Elvis et al. (1994) and Zhang \& Marscher
(1994). O'Dea et al. (1996) report the detection of a luminous
($L_X \sim 2 \times 10^{43}$~erg~s$^{-1}$) and highly absorbed
($N_H \simeq 4 \times 10^{22}$~cm$^{-2}$)
hard X-ray source in 1314+125
($z = 0.122$). Siemiginowska et al. (2002) recently
discovered with {\it Chandra} a prominent
X-ray jet on scales $\sim 300 h^{-1}_{50}$ in PKS~1127-145
($z = 1.187$).

\section{OQ+208: discovery of the first radio-loud Compton-thick AGN}

We present in this paper results
of the January 1999 ASCA observation of the GPS
radio galaxy OQ+208 (1404+286; $z = 0.077$). The
\begin{figure}
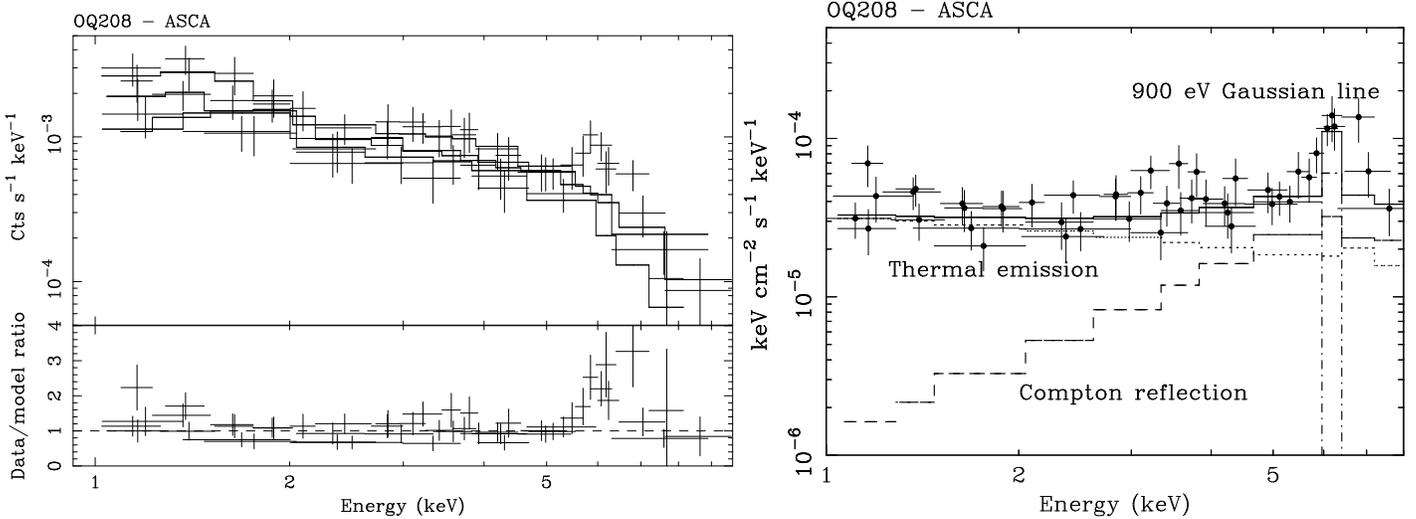

\centerline{\psfig{figure=matteo_guainazzi_fig1.ps,width=9.7cm,angle=-90}
            \psfig{figure=matteo_guainazzi_fig2.ps,width=8.8cm,angle=-90}}
\caption[]{
{\it Left panel:}
{1404+286 ASCA spectra ({\it upper panel}) and residuals in units
of data/ratio models ({\it lower panel}) when a simple
power-law model is applied. Photoelectric absorption
column density due to our Galaxy ($N_H = 1.4 \times 10^{20}$~cm$^{-2}$)
is assumed.
{\it Right panel}: Unfolded best-fit model superimposed to the ASCA
spectra of the same objects in the Compton-reflection plus
thermal plasma scenario. The individual components are
labeled }
}
\label{fig1}
\end{figure}
inverted X-ray spectrum (see Fig.~\ref{fig1};
energy index $\alpha = -0.17 \pm^{0.16}_{0.17}$) and
the remarkably strong
(Equivalent Width, EW, $900 \pm400$~eV)
K$_{\alpha}$ fluorescent 
neutral iron  line ($E_{centroid} = 6.49 \pm 0.09$~keV)
are best
explained if the X-ray emission is dominated by the
Compton-reflection of an otherwise invisible continuum.
The column density covering the AGN is likely to be
as large as $10^{24}$~cm$^{-2}$. If this explanation
is correct, 1404+286 would be {\it the first 
radio-loud Compton-thick AGN ever observed}.
The alternative possibility of a Compton-thin
absorber (still with $N_H \simeq 1.2 \times 10^{23}$~cm$^{-2}$)
is equally viable from the statistical point of view,
although in this scenario there is no clear explanation for the
extreme intensity of the iron line (Leighly \& Creighton 1993).
In both cases, a soft excess below 3\,keV is present,
which can be explained as thermal emission from an
optically thin hot gas, whose temperature is,
however, loosely constrained
($T > 3.8 \times 10^7$~K; cf. the right panel
of Fig.\,\ref{fig1}). Again, the
quality of the ASCA data is not good enough to rule out
alternative explanations for the soft X-ray emission, as
electron scattering of the primary nuclear emission.
The observed fluxes
are 0.7 and $5.5 \times 10^{-13}$~erg~cm$^{-2}$~s$^{-1}$
in the 0.5--2~keV and 2--10~keV bands, respectively.
They correspond
to rest-frame unabsorbed
luminosities of $1.7 \times 10^{42}$ and
$1.4 \times 10^{43}$~erg~s$^{-1}$ in the same
energy bands. However, if the bulk
of the hard X-ray emission is indeed due to Compton-reflection,
the true intrinsic 2--10~keV luminosity of 1404+286 is
likely to be at least one order of magnitude higher
(Guainazzi et al. 1999; Awaki et al. 2000).

Although the measurement of the iron line is robust, and
points unambiguously to a reflection-dominated spectrum,
the overall spectral deconvolution is still
ambiguous. Time has been allocated for the XMM-Newton
2$^{\circ}$ observational cycle (January-September 2003)
to reobserve this target. The unprecedented collecting
power of XMM-Newton in the
hard X-rays  (Jansen et al. 2001)
will allow us to unambiguously characterize all the spectral
components in this intriguing source.

\section{The XEUS perspective}

\begin{figure}
\centerline{\psfig{figure=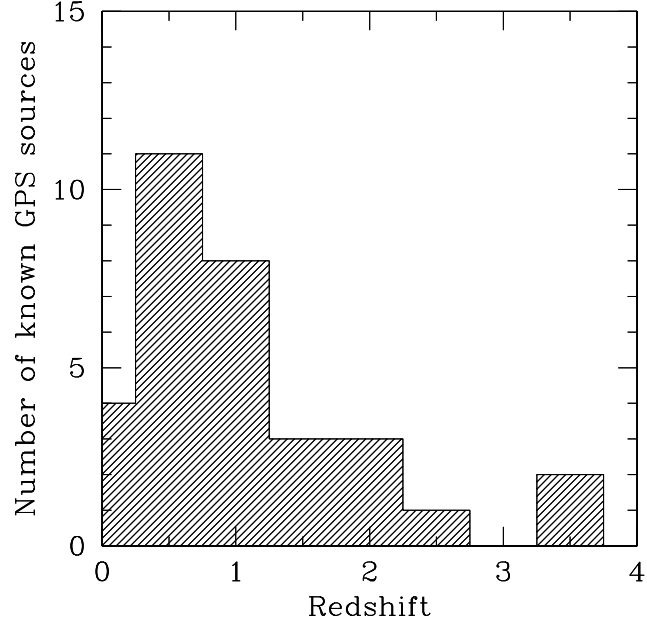,width=8.8cm}}
\caption[]{Redshift distribution of the known GPS
sources after O'Dea 1998.}
\label{fig3}
\end{figure}

\begin{figure}
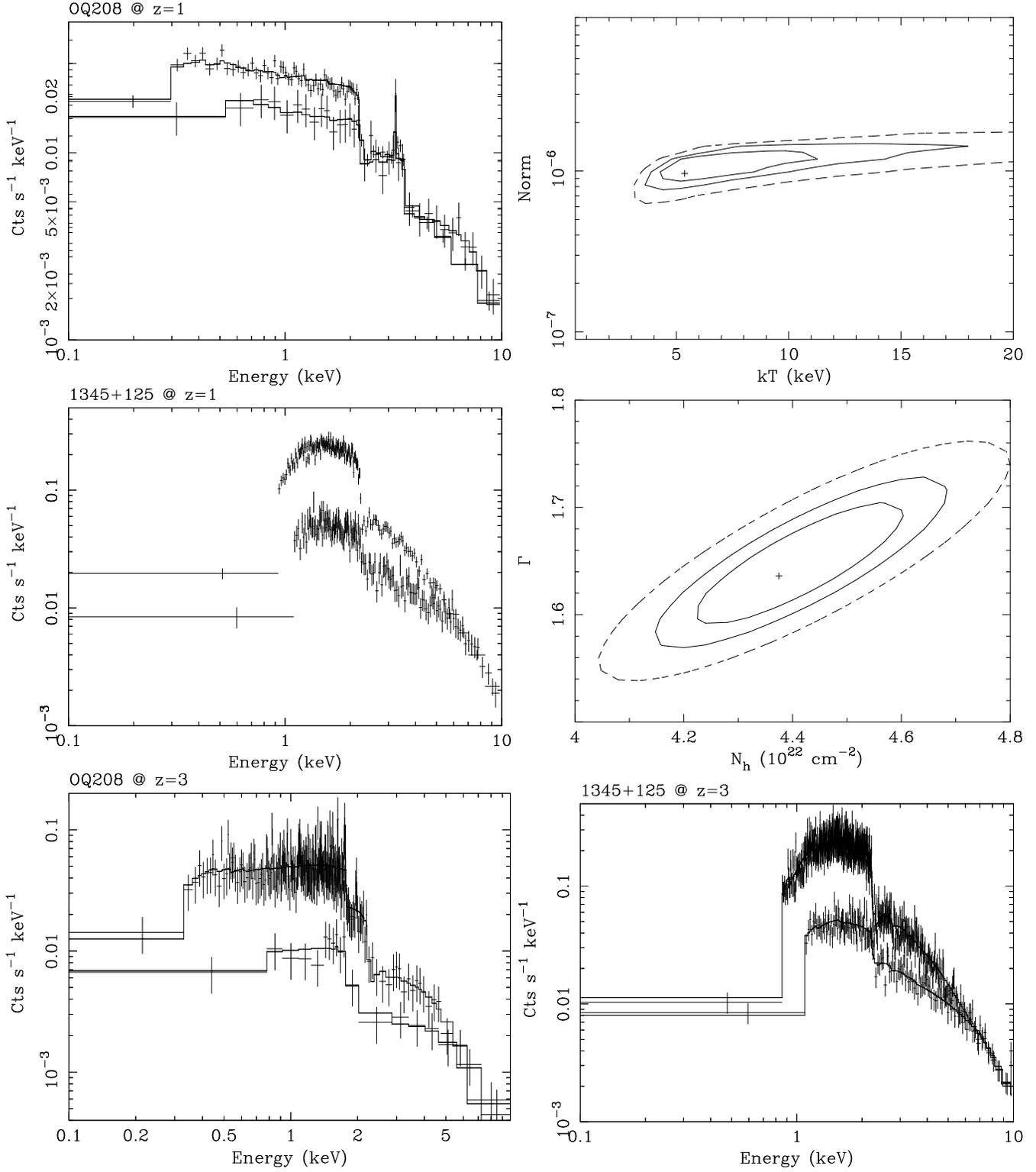

\centerline{\psfig{figure=matteo_guainazzi_fig4.ps,width=8.8cm,angle=-90}
            \psfig{figure=matteo_guainazzi_fig7.ps,width=8.8cm,angle=-90}}
\centerline{\psfig{figure=matteo_guainazzi_fig5.ps,width=8.8cm,angle=-90}
            \psfig{figure=matteo_guainazzi_fig8.ps,width=8.8cm,angle=-90}}
\centerline{\psfig{figure=matteo_guainazzi_fig6.ps,width=8.8cm,angle=-90}
            \psfig{figure=matteo_guainazzi_fig9.ps,width=8.8cm,angle=-90}}
\caption[]{XEUS simulation of GPS sources at cosmological
redshifts. Details in text.}
\label{fig2}
\end{figure}

Only with XEUS one can get spectroscopic information on GPS sources at
cosmological redshifts, where most of them are located
(cf. Fig.~\ref{fig3}).
Simulations have been done, placing 
at redshift of $z = 1$ (Fig.~\ref{fig2};
{\it upper} and {\it central} panels) and
$z = 3$ ({\it lower} panels)
the two GPS
sources for which hard X-ray measurements are available
(see above). 
An exposure time of 50~ks has been assumed throughout,
and the CIS Superconducting Tunnelling Junctions detectors response
matrices employed.
In each spectral plot, the lower curve corresponds to the initial
configuration,
the upper curve corresponds to the final configuration one. The iso-$\chi^2$
contours plots refer to the latter case. 

\end{document}